\documentclass[MSNbibl,number,citesort,dvips]{arxstspdf}
\usepackage{flushend}
\usepackage{stfloats}
\usepackage{graphicx}
\volume{29}
\issue{1}
\pubyear{2014}
\firstpage{42}
\lastpage{49}
\doi{10.1214/13-STS431} 

\begin{document}
\begin{frontmatter}

\title{Experiences in Bayesian Inference in Baltic Salmon Management}
\runtitle{Experiences in Bayesian Inference in Baltic Sea Management}

\begin{aug}
\author[a]{\fnms{Sakari} \snm{Kuikka}\ead[label=e1]{sakari.kuikka@helsinki.fi}},
\author[a]{\fnms{Jarno} \snm{Vanhatalo}\corref{}\ead[label=e2]{jarno.vanhatalo@helsinki.fi}},
\author[b]{\fnms{Henni} \snm{Pulkkinen}\ead[label=e4]{henni.pulkkinen@rktl.fi}},
\author[a]{\fnms{Samu} \snm{M\"antyniemi}\ead[label=e3]{samu.mantyniemi@helsinki.fi}}
\and
\author[c]{\fnms{Jukka} \snm{Corander}\ead[label=e5]{jukka.corander@helsinki.fi}}
\runauthor{S. Kuikka et al.}

\affiliation{University of Helsinki, University of Helsinki, Finnish Game and Fisheries Research Institute, University
of Helsinki and University of Helsinki}

\address[a]{S. Kuikka is Professor, J. Vanhatalo is Post Doctoral Researcher, S. M\"antyniemi is Professor, Department of Environmental Sciences, University of Helsinki,
P.O. Box 65, FI-00014 University of Helsinki, Finland \printead{e1}; \printead*{e2}; \printead*{e3}.}
\address[b]{H. Pulkkinen is Post Graduate Student, Finnish Game and Fisheries Research
Institute, PL 413, 90014 Oulun yliopisto, Finland \printead{e4}.}
\address[c]{J. Corander is Professor, Department of Mathematics and Statistics, University of Helsinki, P.O. Box 68, FI-00014 University of
Helsinki, Finland \printead{e5}.}

\end{aug}

%
\begin{abstract}
We review a success story regarding Bayesian inference in
fisheries management in the Baltic Sea. The management of salmon
fisheries is currently based on the results of a complex Bayesian
population dynamic model, and managers and stakeholders use the
probabilities in their discussions. We also discuss the technical
and human challenges in using Bayesian modeling to give practical
advice to the public and to government officials and suggest
future areas in which it can be applied. In particular, large
databases in fisheries science offer flexible ways to use
hierarchical models to learn the population dynamics parameters
for those by-catch species that do not have similar large
stock-specific data sets like those that exist for many target
species. This information is required if we are to understand the
future ecosystem risks of fisheries.
\end{abstract}

%

\end{frontmatter}
%
\section{Introduction}

We introduce a case of fisheries management where Bayesian inference
has been extensively used. Fisheries management is a field of applied
science, and one could easily argue that fisheries science is as close
to politics as science can be. Fisheries scientists routinely advise
managers and politicians about possible catch allocations for the near
future. This advice has to be concentrated on aspects relevant to the
objectives defined by legislation and international agreements
\cite{Hilborn:2012}. Such advice is a highly charged issue, since
fishing is probably the best known example of the tragedy of commons
(i.e., the depletion of a shared resource by individuals contrary
to the group's\vadjust{\goodbreak} long-term best interests \cite{Hardin:1968}) brought
into public awareness by the collapse of arctic cod stocks in 1992
which rapidly resulted in the loss of over 40,000 jobs in Canada
\cite{Hutchings+Myers:1994}. Even though Bayesian models are becoming
increasingly common in fisheries management due to the adoption of the
precautionary approach, it remains a challenge for a scientist to tell
a fisherman, ``You need to cut down your income this year because I am
so uncertain about the consequences of your fishing.''\looseness=1

\begin{figure*}
\centering
\begin{tabular}{@{}c@{\hspace*{2pt}}c@{}}

\includegraphics{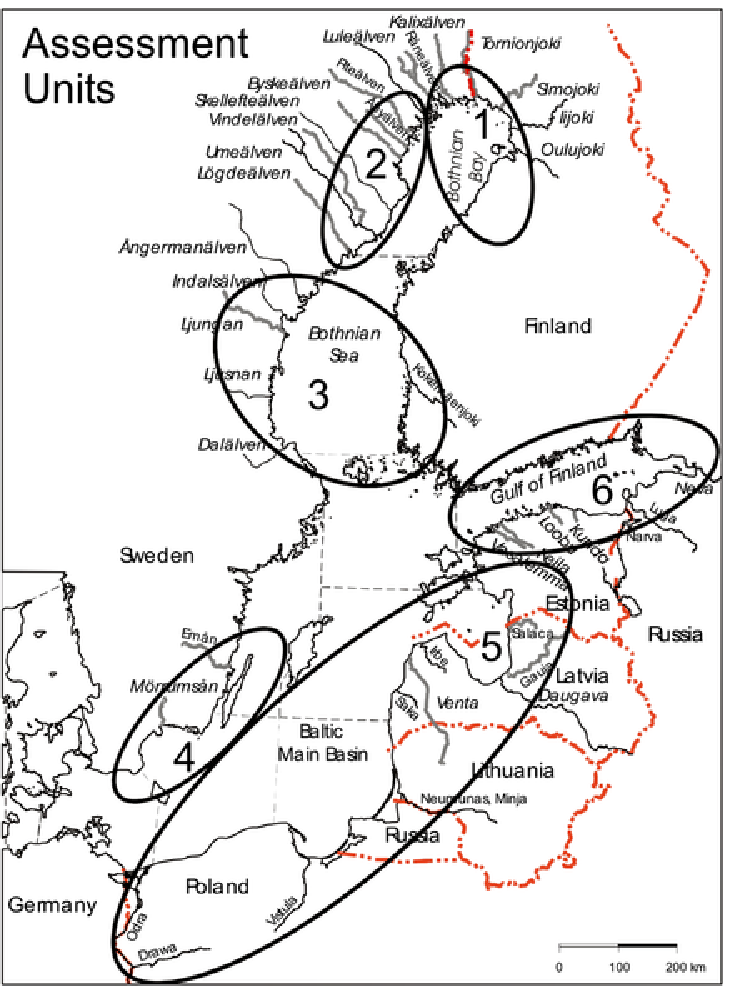}
 & \includegraphics{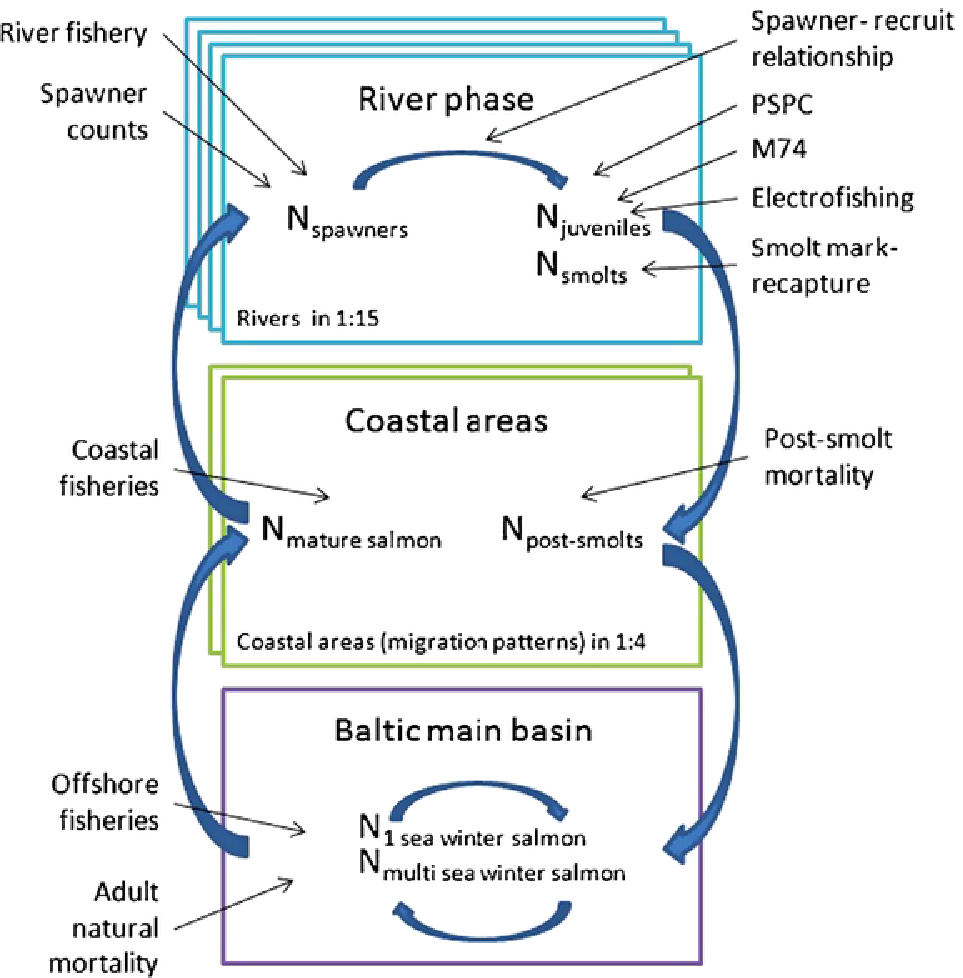}\\
\footnotesize{(a) The spawning rivers and assessment units} &
\footnotesize{(b) The life cycle of
salmon}\\
\footnotesize{of Baltic
salmon \cite{ICES:2011}} &
\end{tabular}
\caption{The Baltic sea and the life cycle of salmon.}\label{fig1}\vspace*{-2pt}
\end{figure*}

Thus, fisheries management is an area of risk analysis where it is
crucial for effective decision-making to utilize all potential
information sources and to make scientifically sound estimates.
Specifically, fish management policymakers need to be given sound
estimates of the uncertainties involved in predictions about how stock
will develop under each alternative management action that can be made
in the near future.

This application requires Bayesian decision analysis~\cite{Kuikka+Hilden+Gislason+Hansson+Sparholt+Varis:1999}, by which
one can analyze the role of alternative information sources in support
of decision-making and the effects of alternative decisions on various
aims of stakeholders and society. Moreover, the possibility of using
expert knowledge in addition to data \cite{Lecklin+Ryoma+kuikka:2011}
is useful when creating complex models for risks which have not yet
occurred.

In addition to the obvious scientific reasons for applying Bayesian
inference in fish stock assessment~\cite{Punt+Hilborn:1997}, there is
also a legislative reason for methodology. Because all fisheries\vadjust{\goodbreak}
legislation has incorporated a precautionary approach \cite{CEC:2009},
policies should be risk averse and account for uncertainty estimates.
By providing scientifically justified statements of uncertainty,
Bayesian stock assessment models can help in such a process. In
particular, assembling prior probabilities from existing literature,
still an underutilized approach, can be useful.\looseness=1

%
%
%

\section{Baltic Salmon Fisheries Management}

\subsection{The Baltic Sea}

The Baltic Sea is a brackish water ecosystem with several unique
features. The salinity varies from around 20 per mille in the south to
close to freshwater at the end of the Bothnian Bay and the Gulf of
Finland [Figure~\ref{fig1}(a)]; as a result, most Baltic sea species
are genetically unique. Predicting future changes in the Baltic sea
ecosystem is challenging owing to, for example, the unpredictable
periods of
low oxygen levels. Future salinity and nutrient levels may also
be different than those observed in historical data
\cite{Fonselius+Valderrama:2003}. It is also expected that climate
change will further impact both the salinity and the temperature of
the sea \cite{Heino+etal:2008}.\vadjust{\goodbreak} For these reasons, historical data
alone is not sufficient for predicting the future.

\subsection{Baltic Salmon}

Baltic salmon are a geographically isolated population of Atlantic
salmon (\emph{Salmo salar} L.), which can be further divided into
subpopulations, called stocks, corresponding to their spawning rivers.
The salmon is a migratory species that spends its first years in a river,
travels to the open sea for its feeding migration and returns to the
river to spawn [see Figure~\ref{fig1}(b)]. Since each salmon
subpopulation returns only to its native river to spawn, maintaining
all stocks in a healthy condition is an important task for fisheries
management. Owing to the high level of exploitation in the early 20th
century, the abundance of wild Baltic salmon dropped significantly
until the 1980s. In addition, the damming of rivers has reduced or
even eliminated the possibility for successful natural reproduction in
many Baltic rivers
\cite
{Romakkaniemi+Pera+Karlsson+Jutila+Carlsson+Pakarinen:2003,ICES:2011,IBSFC+HELCOM:1999}.
In order to compensate for the losses of natural reproduction,
hydropower companies are obliged to release reared salmon annually
into the mouths of dammed rivers. This activity increases the fishing
potential but provides yet another challenge for wild salmon
management, since the fisheries cannot distinguish between reared and
wild salmon. Recruitment of reared stocks cannot collapse due to
overfishing, while many wild stocks have collapsed.

The migration routes of salmon are long, extending from the
northernmost spawning river, the Tornionjoki River, to the feeding
areas in the southern Baltic Sea. These long migration routes expose
the salmon to high pressure from fishing and lead to political debates
about ``who is taking our salmon'' between the coastal countries of
the Baltic Sea. On the other hand, spatial migration offers a lot of
data from various parts of the life cycle of salmon that can be
utilized in population dynamics models applied to the Baltic Sea
salmon assessments.

\subsection{Baltic Salmon Fisheries Management}

The Baltic fisheries are controlled by the EU's Common Fisheries
Policy \cite{CEC:2009} and by bilateral agreements between the EU and
Russia that aim at ensuring economically, environmentally and socially
sustainable fisheries. The management decisions concerning the EU
fisheries are made annually. Based on scientific advisories, the
European Commission prepares proposals for management measures and the
actual regulations are adopted by the Council of Fisheries Ministers.
In 1997, new international long-term management goals were agreed upon
and incorporated into the Salmon Action Plan \cite{IBSFC+HELCOM:1999}.
One of the most important goals was to safeguard the wild salmon
populations by attaining at least 50\% of the potential smolt
production capacity (PSPC) in each wild salmon river by 2010. Smolts
are juvenile salmon that leave their native river for the feeding
migration at the sea. The aim was not easy to implement in practice, as
the stock specific PSPCs were poorly known
\cite{Uusitalo+Kuikka+Romakkaniemi:2005}.

Thus, the Salmon Action Plan created a need to enhance scientific
knowledge about different stages of the salmon life cycle. Because
salmon population dynamics are complex and data about most of the
stocks are sparse, the only realistic approach for developing the
necessary decision tools is Bayesian modeling. Compared to more
traditional statistical methods, Bayesian models make it possible to
combine relevant data from many sources and integrate their
information content with a vast amount of expert knowledge in a
probabilistic manner. Such a framework provides both estimates for the
historical status of the stock and predictions for future stock
development under diverse possible management actions. Thanks to the
Bayesian interpretation of their probabilities, one can also answer
the essential questions of interest, such as, ``What are the
probabilities for each stock reaching 50\% or 75\% of the PSPC in the
next four to six years?'' Fisheries management decisions must have
their desired effect within this time period because salmon stocks
have short life cycles.

\begin{figure*}

\includegraphics{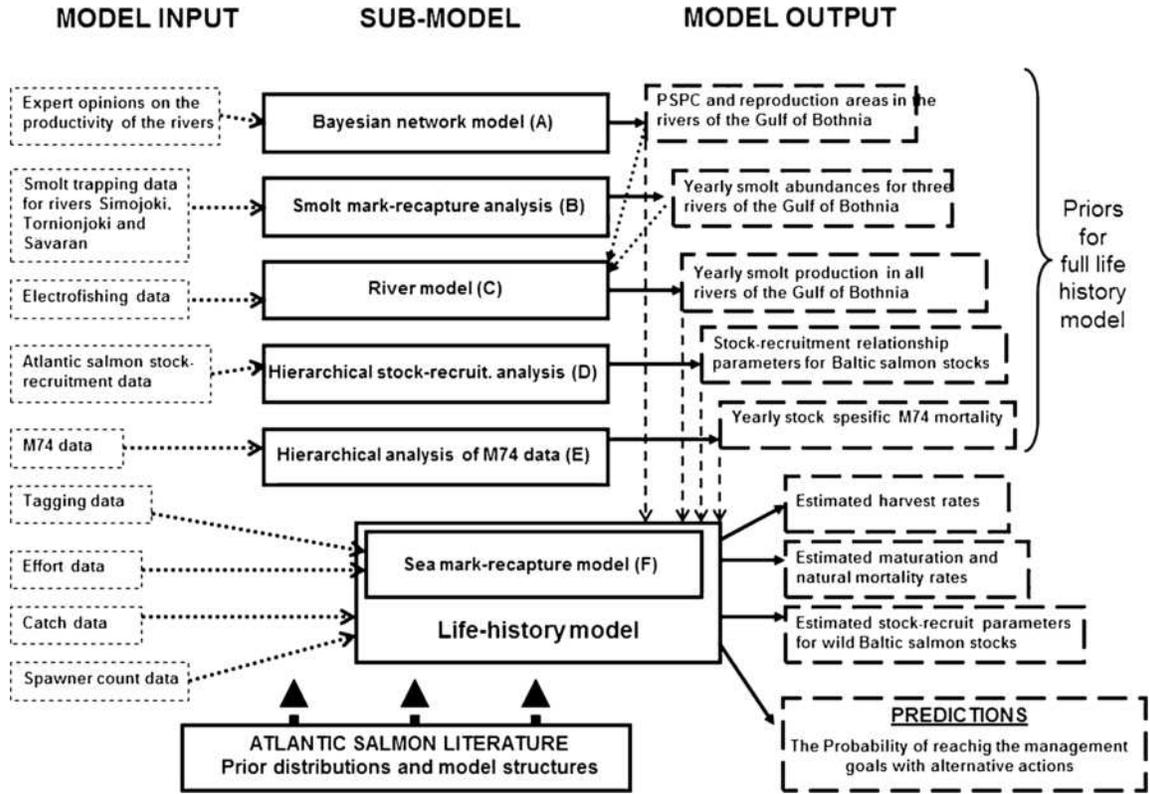}

\caption{The structure of the Baltic salmon assessment model. The
most essential blocks of the model are shown in the boxes
enclosed by solid lines. The data are illustrated with thin
dashed-line boxes on the left and the model outputs with thick
dashed-line boxes on the right~\cite{ICES:2011}.}\label{Figure3}
\end{figure*}

\subsection{Assessment of Baltic Salmon Stocks}

In the beginning of the 1990s, Baltic salmon stock assessment was
performed using simple spreadsheet calculus without any Bayesian
features. One of the problems with such deterministic models is that
natural and fishing mortalities are assumed to be known without
uncertainty, and that the values chosen have a huge impact on the
abundance estimates of the stocks. After Varis and Kuikka
\cite{Varis+Kuikka:1997} first applied Bayesian inference to salmon
assessment in the Baltic Sea, the need to distinguish effectively
between well-known and poorly-known populations led to the wide
application of hierarchical models. Today the scientific advisory on
Baltic salmon is entirely based on Bayesian methods. The greatest
advantage of Bayesian models compared to the traditional statistical
models is that uncertainties in large data sets with lots of variation
will be taken into account and be visible in the posterior estimates.
Thus, traditional methods that are based on point estimates are
considered misleading and dangerous by fisheries scientists familiar
with the current methods. When scientific advice is given for the
management of the stocks, it is highly important to take into
consideration the probability of not reaching management objectives
because of certain fishing pressure. This is possible only within
Bayesian models.\looseness=-1

The objectives of wild salmon fisheries management include ensuring a
level of smolt production in rivers that will keep the stocks alive
and healthy and, at the same time, enable salmon fisheries. Thus, the
main focus of the stock assessment is to predict the near-future
development of stocks under alternative management plans. However, in
order to undertake this, it is important to acknowledge the high
complexity of the salmon life cycle and to model all of the factors
that influence salmon survival at different life stages in a
biologically justifiable manner. Only by understanding the reasons
behind the historic changes in levels of abundance and the
uncertainties in the biological processes is it possible to advocate
management actions that will both prevent the stocks from collapsing
and enable their sensible economic exploitation.

Current salmon stock-assessments are based on describing the
population life history using an age-structured state--space model (see
life-history model, Figure~\ref{Figure3})
\cite
{Michielsens+McAllister+Kuikka+Mantyniemi+Romakkaniemi+Pakarinen+Karlsson+Uusitalo:2008}.
The state variables describe the temporal and spatial changes in the
demography of the salmon population. These include the abundance of
wild smolts, $R_{i,t}$, the abundance of salmon in the sea,
$N_{i,t,a}$, the spawning population, $S_{i,t,a}$, and the number of
eggs, $O_{i,t}$, for each stock $i$ and year $t$. The subscript $a$
denotes the number of years the salmon have spent in the sea after
leaving the river. The model structure and state transitions are
described according to existing biological knowledge about the life
cycle of salmon, which is illustrated in Figure~\ref{fig1}(b). For
example, the transition from smolts to the one-sea-year salmon
population is controlled by the general relation, $N_{i,t+1,1} =
R_{i,t}\exp(-F_{t,0}-M_{t,0})\varepsilon$, where $F$ and $M$ are the
instantaneous fishing and natural mortality rates and $\varepsilon$ is
the process error
\cite
{Michielsens+McAllister+Kuikka+Pakarinen+Karlsson+Romakkaniemi+Pera+Mantyniemi:2006}.

The number of eggs produced by stock $i$ is linearly dependent on the
stock's spawning population,
$S_{i,t,a}=L_aN_{i,t,a}\exp(-F_{t,a}-M_{t,a})\varepsilon$, where $L_a$ is
the fraction of the salmon population maturing at sea year age $a$.
The recruitment of new smolts is described by the Beverton--Holt
\cite{Beverton+Holt:1957} stock-recruit (SR) function $R_{i,t+T} =
O_{i,t}/(\alpha_i + \beta_iO_{i,t})$, which describes the relationship
between the number of eggs and the abundance of new recruits $T$ time
steps later. This function and its parameter values are some of the
most important model specifications in fisheries stock assessment,
since they determine the predicted impacts of management actions on
stock development in the future \cite{Michielsens+McAllister:2004}.
Another important factor in Baltic salmon recruitment is early
mortality syndrome, M74 syndrome, outbursts of which can kill the
majority of juveniles during their first year of life
\cite{Michielsens+Mantyniemi+Vuorinen:2006}.

The inference for the life-history model is performed sequentially, as
illustrated in Figure~\ref{Figure3} and described by Michielsens
et~al.
\cite
{Michielsens+McAllister+Kuikka+Mantyniemi+Romakkaniemi+Pakarinen+Karlsson+Uusitalo:2008}.
The Bayesian models A, D, E provide prior distributions for the
parameters of the life-history model. These prior distributions are
based on expert knowledge (e.g., PSPC, A), data from other Atlantic
salmon stocks (e.g., parameters of the stock-recruit function, D~\cite{Michielsens+McAllister:2004}) or data sets of Baltic salmon that
are computationally too heavy to analyze within the full life-history
model but can be analyzed separately (e.g., the early mortality
syndrome M74 model, E). The posterior distributions of parameters from
these models are used as informative prior distributions in the
life-history model. The (final) posterior distributions of parameters
and state variables are calculated by conditioning on (indirect)
observations of the state variables. These data include the time
series of catch and effort from different fisheries, spawner count
data sets for some rivers, Carlin-tag mark recapture data and data
about smolt abundances for a number of rivers. All of the data sets
except the last one directly provide likelihood functions for the
state variables. Since the observation model for the annual smolt
abundances is computationally too demanding, it is approximated as
described below.

\begin{figure}

\includegraphics{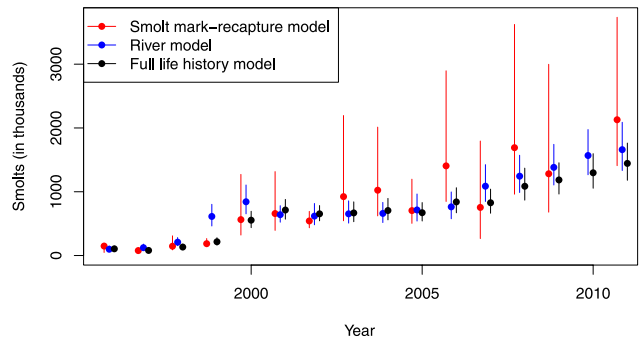}

\caption{The estimate of smolt abundance in the Tornionjoki River
after sequential modeling steps B, C and the life-history model
shown in Figure \protect\ref{Figure3}.}\label{TornionjoenSmoltit}
\end{figure}

The data for smolt abundances consist of river-specific smolt
mark-recapture data and electrofishing data that contain information
about parr (juvenile, 1--4 year-old salmon) densities. The
electrofishing data are more often available than the smolt
mark-recapture data. First a mark-recapture analysis (model B,
\cite{Mantyniemi+Romakkaniemi:2002}) is conducted for rivers with
mark-recapture data. These results are then used in a river model
(model C, \cite{Mantyniemi+Romakkaniemi+Uusitalo+Arjas:2004}) which
describes the relationship between the smolt and parr abundances. The
river model is hierarchical over all rivers and, thus, provides smolt
abundance estimates for all wild salmon stocks. The posterior
distributions of yearly smolt abundances provided by model C are used
to construct an approximation for the likelihood function with respect
to $R$ in the life-history model~\cite
{Michielsens+McAllister+Kuikka+Mantyniemi+Romakkaniemi+Pakarinen+Karlsson+Uusitalo:2008}.

As an example of updating the parameter estimates in the
sequential model framework, we illustrate the case of the annual smolt
abundance estimates for Tornionjoki River salmon, which is one of the
few rivers for which smolt mark-recapture data exist
\cite{Mantyniemi+Romakkaniemi:2002}. Figure~\ref{TornionjoenSmoltit}
shows how the estimate of smolt abundance changes in each successive
modeling step. The posterior uncertainty is highest after modeling
step B, but as data accumulates the posterior distribution becomes
tighter. Ideally, we could infer all the models in Figure~\ref{Figure3} jointly, but this is not possible with the current
computational tools within a reasonable time frame.

\section{Discussion}

\subsection{Why Bayes?}

The management problem highlighted above is an example of a problem
which could not have been solved efficiently without Bayesian methods.
Here we summarize the most important reasons for using Bayes:

\begin{itemize}
\item The decision problem is multidimensional. There are several
stakeholders with different aims and, thus, the statistical methods
used have to allow a detailed decision analysis.
\item The life-cycle of Baltic salmon and its response to natural and
human induced pressures are complex. Thus, in order to take all the
plausible uncertainties into account in decision-making, the stock
assessment model needs to reflect the biological realism. This leads
to a model with so many parameters that they cannot be estimated
without the use of informative priors.
\item The available data are multifaceted and there is available
essential prior knowledge complementary to data. Thus, the
statistical methods must allow hierarchical model structures and the
explicit use of priors.
\item The precautionary approach incorporated in EU fisheries
legislative demands for methods that take explicitly into account
all sources of uncertainty.
\end{itemize}
%

\subsection{Future Scientific Challenges}

The greatest current challenges involved in the above example are
twofold. First, the computational and technical tasks related to
Bayesian inference are complex and time-consuming. Second, it is
necessary that the people involved in modeling have a sufficient
subject understanding. The selection of model structures, prior
probabilities and the likelihood function(s) all depend on subject
matter knowledge, in this case biological knowledge. However,
researcher's subject matter background easily means that computational
problems become overwhelming. It seems that educating methodologically
orientated scientists in biology is relatively an easier task than
educating, for example, biologists in Bayesian inference.

Although fairly easy-to-use software is available (e.g., OpenBugs or
JAGS), much of the time spent by biologically trained scientists is
allotted to technical problems related to MCMC algorithms and in
waiting for the convergence of runs. This does not represent optimal
use of scientific resources. Moreover, because the final modeling is
usually based on the outcomes of earlier analysis, and is the last
step of a big project, computational problems easily lead to failure
in timing.

The Bayesian approach offers a way to formalize scientific learning.
The posterior distributions of one study can be used as priors in
following studies, if the results are published in a transparent and
sufficient way. In fisheries, the risk analyses needed for by-catch
species in particular are more or less impossible without
meta-analysis \cite{Myers+Mertz:1998}. It is necessary to learn more
effectively from existing databases and publications and to apply,
for example, hierarchical Bayesian models to provide informative priors for
case-specific risk analyses and to utilize the correlations of
biological features of species in order to make better predictions
\cite{Kuparinen+etal:2012,Pulkkinen+mantyniemi+kuikka+levontin:2011}.

The scientific tradition of publishing only ``statistically
significant'' results is a major problem for meta-analysis and
systematic learning processes. If only extreme data sets (say, with
$p< 0.05$) are published and used in meta-analyses or in scientific
discussions, a biased view of the system can be easily obtained. On
the other hand, using published papers as a source of prior
information in Bayesian models can also create problems when, for
example, it
is uncertain whether published values are representative samples of
the system studied.

The allocation of resources used for data analysis or, alternatively,
for the analysis of priors should be an interactive process during
scientific projects. If the available data will not be informative
enough to make justified scientific and management conclusions,
major effort should be directed toward effective and justified
derivation of priors. In some cases, this may even be a longer process
than a ``traditional'' data analysis. The collection of new data can be
very costly compared to the use of published papers or existing
databases \cite{Pulkkinen+mantyniemi+kuikka+levontin:2011}.

Sometimes it may either be very expensive or difficult to collect data
about variables of interest. In such cases knowledge must be elicited
from experts. While frequentist methods could be used to (point)
estimate some of the model parameters, large parts of the system would
be entirely left out from the quantitative analysis owing to complete
lack of data. Moreover, Uusitalo et~al. have demonstrated
\cite{Uusitalo+Kuikka+Romakkaniemi:2005} that the highest uncertainty
in expert knowledge related to salmon assessment comes from the fact
that expert opinions differ, and Bayesian inference is needed to
integrate those sources of uncertainty. Thus, the classical approach
could not provide appropriate answers to the problem of management
under uncertainty.

The multifaceted nature of the salmon assessment problem requires use
of a complex model, easily leading to thousands of unknown parameters.
A vast amount of data with high spatio-temporal resolution would be
required to sufficiently identify all of these parameters, if point
estimation without prior knowledge was desired. A reasonable maximum
likelihood estimation of the main target parameters would require
reduction of the model dimensions by effectively assuming that many of
the uncertain nuisance parameters were actually known
\cite{Kuparinen+etal:2012}. From the decision-making point of view,
this implies that management would then be based on overconfident
estimates. While not easy to conduct, the Bayesian approach has made
it technically possible to attempt realistic stock assessment, which
would currently not be feasible with any other methods.

\subsection{Challenges in Applying Bayesian Inference to Practical
Scientific Advice}

As mentioned earlier, fisheries science is very close to political
decision-making and, as in any attempts to model complex systems,
there are many subjective choices involved before model-based advice
can be given. However, it is far easier for a scientist to defend an
analysis when as much data and as few obviously subjective choices as
possible have been included. The time available during meetings of
stock \mbox{assessment} working groups is often too limited for complex
Bayesian models to be applied during the meeting, as the computational
inference may easily require a week or more to converge. Thus,
transition to Bayesian methods would also demand changes in the
practices of the working group in a way that part of the work would
need to be done beforehand instead of at the last minute of the
meeting.

The need for faster algorithms is obvious, since understanding of the
model dynamics and of the logic by which the model operates require
practically short calculation time to allow for ``what if'' type of
questions. It is common that in the working groups of fisheries stock
assessment the latest data (most recent year on which the predictions
are based) can give rise to surprising results. These need then to be
discussed and the models adapted accordingly during a very short
period of time if the results are to be explained to representatives
of the industry and other stakeholders. In some cases this implies
that the sensitivity of modeling outputs has to be tested against
alternative informative priors. Such sensitivity analyses are
critically needed to explain the behavior of complex models to the end
users of the information in order to improve their commitment to
modeling results
\cite{Haapasaari+Michielsens+Karjalainen+Reinikainen+Kuikka:2007}.
Unfortunately, fisheries scientists are not optimally trusted
\cite{Glenn+etal:2011} by industry and improvements in this regard
require fully open approaches and learning improved ways to
communicate risks. Salmon assessment models are by necessity complex
given the characteristics of the life cycle of fishes, but a central
goal is nevertheless to make the results more easily understood. When
inference algorithms are so slow that only a single run is possible
during a week-long working group meeting, the model behavior may not
be understood well enough. One approximate solution to this, as is
done, for example, in oil spill risk analysis, is to use estimation models
and feed the posterior information to a Bayesian network inference
engine, which allows an interactive use of the probabilistic results,
albeit only approximately
\cite
{Kuikka+Hilden+Gislason+Hansson+Sparholt+Varis:1999,Levontin+Kulmala+Haapasaari+Kuikka:2011}.

To facilitate general adoption of Bayesian reasoning in risk-averse
decision-making, scientists must try to broaden public understanding
about risks and the projected consequences of different policies, in a
way that is similar to the ongoing debate around climate change. We
call for experts in the cognitive sciences to test systematically how
uncertainties should be communicated. Such developments will help
prevent more manmade fisheries catastrophes such as the arctic cod
stock collapse of 1992. Since aquatic resources are in global decline
and the situation is already alarming for many ecologically and
economically important species, there is more need now than ever for
careful Bayesian reasoning to help improve the public's understanding
of the risks facing these resources.

\section*{Acknowledgments}

This work is partly supported by the Academy of Finland (Finnish
Centre of Excellence in Computational Inference Research COIN,
251170), the Natural Sciences and Engineering Research Council of
Canada (JH) and from the European Union's Seventh Framework Programme
(FP7/ 2007-2013) under Grant agreement No.~244706/ECOKNOWS project (SK,
SM, JV). However, this paper does not necessarily reflect the European
Commission's views and in no way anticipates the Commission's future
policy in the area.

%



\end{document}